\documentclass[prd,twocolumn,aps,amsmath,amssymb,nofootinbib,preprintnumbers]
{revtex4}
\IfFileExists{srcltx.sty}{\usepackage[active]{srcltx}}%

\usepackage{graphicx}% Include figure files
\usepackage{dcolumn}% Align table columns on decimal point
\usepackage{bm}% bold math
\usepackage{amsmath}
\usepackage{amsfonts}
\def\ls{\mathrel{\lower4pt\vbox{\lineskip=0pt\baselineskip=0pt
           \hbox{$<$}\hbox{$\sim$}}}}
\def\gs{\mathrel{\lower4pt\vbox{\lineskip=0pt\baselineskip=0pt
           \hbox{$>$}\hbox{$\sim$}}}}
\def\drawbox#1#2{\hrule height#2pt

\hbox{\vrule width#2pt height#1pt \kern#1pt
              \vrule width#2pt}
              \hrule height#2pt}

\def\Asym#1#2{\vcenter{\vbox{\drawbox{#1}{#2}
              \kern-#2pt       % line up boxes
              \drawbox{#1}{#2}}}}

\newcommand{\beq}{\begin{equation}}
\newcommand{\eeq}{\end{equation}}

\begin{document}

\title{Spectral tilt in A-term inflation}

\author{Rouzbeh Allahverdi$^{1}$}
\author{Anupam Mazumdar$^{2}$}

\affiliation{$^{1}$~Perimeter Institute for Theoretical Physics, Waterloo, ON, 
N2L 2Y5, Canada \\
$^{2}$~NORDITA, Blegdamsvej-17, Copenhagen-2100, Denmark}

\begin{abstract}
Recently in hep-ph/0605035 and hep-ph/0608138, we have shown that
primordial inflation can be embedded within the Minimal Supersymmetric
Standard Model, while providing the right amplitude for the density
perturbations and a tilted spectrum which matches the current data.
In this short note we show that the model predicts a range of spectral
tilt, $0.92 \leq n_s \leq 1$, depending on deviation from the saddle
point condition. The spectral tilt saturates the lower limit when the
saddle point condition is met. On the other hand the upper limit can
be achieved for a slight deviation towards the point of
inflection. The running of the spectral tilt remains small, and the
amplitude of the temperature anisotropies remains in the correct
observational regime.
\end{abstract}
\maketitle

%%%%%%%%%%%%%%%%%%%%%%%%%%%%%%%%%%%%%%%%%%%%%%%%%%%%%%%%%%%%%%%%%%%%%%%%%%%%%%%

\section{Introduction}

Recently in two Refs.~\cite{AEGM,AKM}, we demonstrated that the
Minimal Supersymmetric Standard Model (MSSM) has all the ingredients
to give rise to a successful inflation (see also~\cite{GMSB}).  Interestingly 
the inflaton belongs to the MSSM, i.e. the flat directions of the
MSSM~\footnote{MSSM has many gauge invariant flat directions made up
of squarks and sleptons, which preserve $R$-parity at the
renormalizable level, for a review see~\cite{MSSM-REV}.}.  In
\cite{AEGM}, the inflaton candidates were the two flat directions,
$LLe$ and $udd$, where $L$ stands for the left-handed sleptons, $e$,
the right-handed sleptons and, $u,~d$ stand for the right handed
squarks (of the up- and down-type respectively).

On the other hand in Ref.~\cite{AKM}, the inflaton was $NH_uL$, the
right handed sneutrino, Higgs and the left handed slepton,
respectively. This is an extension of the MSSM which includes Dirac
neutrinos with masses in the correct range to explain the atmospheric
neutrino anomaly.

The difference between the two models is; that both $LLe$ and $udd$
flat directions are lifted by the non-renormalizable superpotential
terms (of order $n=6$) which are suppressed by the Planck scale,
$M_{\rm P}$, while $NH_uL$ is lifted by the tiny neutrino Yukawa
couplings.

There are many advantages of the MSSM inflation, first of all the
inflaton is not an {\it ad-hoc} gauge singlet from the hidden sector
as often assumed in the literature~\footnote{See Ref.~\cite{JM} for
other attempts of embedding inflation with gauge invariant SUSY flat
directions, where multiple flat directions assisting inflation similar
to that in Refs.~\cite{ASSIST}.}.  Secondly, the inflaton couplings to
the MS(SM) are known, therefore, reheating and thermalization can be
understood consistently within the framework of
supersymmetry~\cite{AVERDI1}. The final reheat temperature is
sufficiently low to avoid any problem with thermal and/or non-thermal
gravitinos~\cite{MAROTO}. The model is insensitive to supergravity
corrections and also does not rely on super-Planckian
VEVs~\cite{AEGJM}. The moduli problem can also be avoided all
together, for all the detailed discussion, see~\cite{AEGJM}.

Although, the scale of inflation is relatively small, $H_{inf}\sim
1$~GeV, the model produces remarkably large number of e-foldings of
inflation. The slow roll inflation is preceded by a phase of inflation
where self-reproduction dominates due to quantum
fluctuations~\cite{LINDE}. During the slow roll, the total number of
e-foldings is found to be ${\cal N}_e\sim 1000$. Not all the
e-foldings of inflation is required to explain the temperature
anisotropies seen in the sky~\cite{WMAP3}, the only relevant number of
e-foldings, which normalizes the COBE amplitude is nearly ${\cal
N}_{\rm COBE}\sim 50$ in our case~\cite{AEGM,AKM}. The details of this
finding depends on thermal history of the universe, see for
instance~\cite{LEACH}.

In this short note, we wish to highlight that the model properties
have become even more richer, especially, the spectral tilt is bounded
from below and above if the predictions of inflation are maintained.
The range of spectral tilt is given by
\beq
0.92\leq n_s\leq 1\,.
\eeq
The lower limit is saturated for an exact saddle point behavior, while
the upper limit is allowed as long as sufficient inflation is
obtained, which is of the order of ${\cal N}_{\rm COBE}\sim 50$ in our
case.

In section 2, we recall some of the important results of a saddle
point inflation, in section 3, we describe a slight deviation from the
saddle point and discuss the spectral tilt and the running of the
spectrum. We also comment on the amplitude of the scalar metric
perturbations.

%%%%%%%%%%%%%%%%%%%%%%%%%%%%%%%%%%%%%%%%%%%%%%%%%%%%%%%%%%%%%

\section{Saddle point inflation}

Let us consider the potential in general, for the radial part of a
flat direction, $\phi$,
\beq \label{scalpot} 
V = {1 \over 2} m^2_{\phi} \phi^2 - A {\lambda_n
\phi^n \over n M^{n-3}_{\rm P}} + \lambda^2_n {\phi^{2(n-1)} \over
M^{2(n-3)}_{\rm P}}\,, 
\eeq
where $A$ is a positive number, $m_{\phi}\sim {\cal O}(1)$~TeV,
$\lambda_{n}\sim {\cal O}(1)$. Within MSSM all the flat directions are
lifted by the non-renormalizable operator, $n \leq 9$. The renormalizable
potential is denoted by $n=3$. The flat direction inflaton $NH_{u}L$
belongs to $n=3$, while $LLe,~udd$ correspond to $n=6$.

When the condition~\footnote{ The importance of $A$-term was first
highlighted in Ref.~\cite{CURV} in connection to inflation and density
perturbations.}
\beq \label{sadcond}
A = \sqrt{8(n-1)} m_{\phi}\,,
\eeq
is satisfied, then there exists a saddle point at
\beq \label{vev}
\phi_0 = \Big({m_{\phi} M^{n-3} \over \lambda_n \sqrt{2n-2}}\Big)^{1/(n-2)}\,,
\eeq
such that $V^{\prime}(\phi_0) = V^{\prime \prime}(\phi_0) = 0$. At
this point we have
\beq \label{pot}
V(\phi_0) = {(n-2)^2 \over 2n(n-1)} m^2_{\phi} \phi^2_0\,,
\eeq
and in its vicinity
\beq \label{sadapr}
V(\phi) = V(\phi_0) + \Big({1 \over 3!}\Big) V^{\prime \prime \prime}
(\phi_0) 
(\phi - \phi_0)^3 + ...,
\eeq
where
\beq \label{3rd} 
V^{\prime \prime \prime}(\phi_0) = 2 (n-2)^2 {m^2_{\phi} \over \phi_0}\,.
\eeq
Hence there is a plateau where the potential is very flat and
inflation is driven by $V^{\prime \prime \prime}(\phi_0)$.

The Hubble expansion rate during inflation is given by
\beq \label{hubble}
H_{\rm inf} = {(n-2) \over \sqrt{6n(n-1)}} {m_{\phi} \phi_0 \over M_{\rm P}}\,.
\eeq
When $\phi$ is very close to $\phi_0$ the first derivative is
extremely small, and we are in a self-reproduction (or eternal
inflation) regime where quantum diffusion is dominant. But eventually
classical friction wins and slow roll begins at $\phi \approx
\phi_{\rm self}$,
\beq \label{self}
(\phi_0 - \phi_{\rm self}) \simeq \Big({m_{\phi} \phi^2_0 \over 
M^2_{\rm P}}\Big)^{1/2} \phi_0\,.
\eeq
The equation of motion for the $\phi$ field in the slow roll 
approximation is given by:
\beq \label{slow}
3 H_{\rm inf} {\dot \phi} = -{1 \over 2} V^{\prime \prime \prime}(\phi_0) 
(\phi - \phi_0)^2\,.
\eeq
Inflation ends when either of the slow roll parameters, $\epsilon
\equiv (M^2_{\rm P}/2)(V^{\prime}/V)^2$ or $\eta \equiv M^2_{\rm
P}(V^{\prime \prime}/V)$, becomes ${\cal O}(1)$. It happens that
$\vert \eta \vert \sim 1$ at $\phi_{\rm end}$, where
\beq \label{end}
(\phi_0-\phi_{\rm end}) \sim {\phi^3_0 \over 4n(n-1)M^2_{\rm P}}\,.
\eeq
We can estimate the total number of e-foldings during the slow roll
phase, from $\phi$ to $\phi_{\rm end}$, 
\beq \label{efold}
{\cal N}_e(\phi) = \int_{\phi}^{\phi_{\rm end}} {H_{\rm inf} d\phi 
\over \dot\phi} \simeq {\phi^3_0 \over 2n(n-1)M^2_{\rm P}(\phi_0 - \phi)}\,,
\eeq
where we have used Eq.~(\ref{slow}). The total number of e-foldings in
the slow roll regime is then found from Eq.(\ref{self}),
\beq \label{tot}
{\cal N}_{\rm tot} \simeq {1 \over 2n(n-1)} \big({\phi^2_0 \over m_{\phi} 
M_{\rm P}}\Big)^{1/2}\,.
\eeq
The observationally relevant perturbations are generated when $\phi
\approx \phi_{\rm COBE}$. The number of e-foldings between $\phi_{\rm
COBE}$ and $\phi_{\rm end}$, denoted by ${\cal N}_{\rm COBE}$, follows
from Eq.~(\ref{efold})
\beq \label{cobe}
{\cal N}_{\rm COBE} \simeq {\phi^3_0 \over 2n(n-1)M^2_{\rm P}(\phi_0 - 
\phi_{\rm COBE})}\,.
\eeq
The amplitude of the scalar perturbations, generated during the slow
roll phase is given by~\cite{AEGM,AKM,AEGJM}:
\beq \label{ampl}
\delta_{H} \equiv \frac{1}{5\pi}\frac{H^2_{\rm inf}}{\dot\phi} \simeq
\frac{1}{5\pi} \sqrt{\frac{2}{3}n(n-1)}(n-2) ~ \Big({m_\phi M_{\rm P} \over 
\phi_0^2}\Big) ~ {\cal N}_{\rm COBE}^2\,,
\eeq    
where we have used Eqs.(\ref{sadapr},\ref{hubble},\ref{cobe}). 

Again after using these equations, the spectral tilt of the power
spectrum and its running are found to be~\cite{AEGM,AKM}
\begin{eqnarray} \label{tilt}
&&n_s = 1 + 2\eta - 6\epsilon \simeq 1 -
{4\over {\cal N}_{\rm COBE}}, \, \\
\label{runtilt}
&&{d\,n_s\over d\ln k} = - {4\over {\cal N}_{\rm COBE}^2}. \,
\end{eqnarray}

For soft supersymmetry breaking parameters $m_{\phi}$ and $A$ in the
range of $1-10$ TeV, perturbations of the correct size are obtained
for $\phi_0 \sim 10^{14}-10^{15}$ GeV~\cite{AEGM,AEGJM,AKM}. This
results in $V(\phi_0) \leq 10^{38}({\rm GeV})^4$. Since reheating in
our case happens instantaneously~\cite{AEGJM}, we find ${\cal N}_{\rm
COBE} \leq 50$~\cite{LEACH}.

The expression for the tilt in the power spectrum, see
Eq.~(\ref{tilt}), then implies that $n_s \simeq 0.92$. Although, the
tilt is compatible with the current WMAP 3-years data within $2
\sigma$~\cite{WMAP3}, it is still somewhat towards the lower side.

A very natural question which arises out of this scenario is whether
the spectral tilt can at all be improved from $0.92$ or not. Note that
the tilt in Eq.~(\ref{tilt}) is a robust prediction of a slow roll
inflation near the saddle point, as it {\it does not} depend on the
detailed form of the potential. Hence any improvement, $n_s > 0.92$
requires deviations from the saddle point condition
Eq.~(\ref{sadcond}).

In the coming section, we argue that it is possible to achieve the
spectral tilt $n_s > 0.92$ within our setup.

%%%%%%%%%%%%%%%%%%%%%%%%%%%%%%%%%%%%%%%%%%%%%%%%%%%%%%%%%%%%%%%%%%%%%%%%%%%%%%
\section{Deviation from the saddle point}

To facilitate the discussion, let us define
\beq 
\label{delta}
\delta \equiv {A^2 \over 8(n-1) m^2_{\phi}} \equiv 1 \pm 
\Big({n-2 \over 2}\Big)^2 \alpha^2,
\eeq
where $\alpha \ll 1$. Note that when $\alpha =0$ and $\delta =1$ we
are back to the saddle point condition~\footnote{A similar analysis is
presented in Ref.~\cite{AEGJM}, but for a different purpose.}.

When $\delta > 1$, we will have a point of inflection at $\phi_0$ and
two extrema (one maximum, one minimum) near it. The field which is
initially trapped in the minimum can tunnel to a point beyond the
maximum {\it a la} Coleman-De Luccia, and a period of slow roll
inflation with a number of e-foldings $\geq {\cal N}_{\rm COBE}$ with
$n_s = 1 - 4/{\cal }_{\rm COBE}$ will follow, provided that $\alpha$
is sufficiently small (for more details
see~\cite{AEGJM}).  For larger values of $\alpha$ results deviate from
the saddle point calculations, which will only result in $n_s <
0.92$~\cite{LYTH1}. Note however that this is not of interest to us as
$n_s$ will lie completely outside the $2 \sigma$ limit from the WMAP
central value~\cite{WMAP3}.

In an opposite case, when $\delta < 1$, there exists only a point of
inflection at $\phi_0$ (i.e. $V^{\prime \prime}(\phi_0) = 0$) where
\beq \label{1st}
V^{\prime}(\phi_0) = \Big({n-2 \over 2}\Big)^2 \alpha^2 m^2_{\phi} 
\phi_0\,,
\eeq 
and $V^{\prime \prime \prime}(\phi_0)$ is given by Eq.~(\ref{3rd}). We
therefore have
\beq \label{1st2}
V^{\prime}(\phi) \simeq V^{\prime}(\phi_0) + {1 \over 2} 
V^{\prime \prime \prime}(\phi_0) (\phi - \phi_0)^2\,.
\eeq
Note that both terms on the right-hand side are positive.  The fact
that $V^{\prime}(\phi_0) \neq 0$ can lead to interesting changes
from a saddle point behavior.

Further note that in the previous section, the slow roll inflation was
driven by $V^{\prime \prime \prime}(\phi_0)$. This holds true only if
the first term on the right-hand side of Eq.~(\ref{1st2}) is
subdominant for $\phi \leq \phi_{\rm COBE}$. After using
Eq.~(\ref{cobe}) this leads to a following bound on parameter
$\alpha$~\footnote{This is the same as the condition for a successful
inflation when $\delta > 1$~\cite{AEGJM}.}
\beq \label{cond1}     
\alpha \ls {1 \over n (n-1) {\cal N}_{\rm COBE}} 
\Big({\phi_0 \over M_{\rm P}}\Big)^2\,.
\eeq
If this bound is satisfied, the spectral tilt will still be given by
Eq.~(\ref{tilt}) and, therefore $n_s \simeq 0.92$ remains valid. In
passing we note that there will be no self-reproduction regime unless
$\alpha \ll (m_{\phi} \phi^2_0/M^3_{\rm P})^{1/2}$~\cite{LYTH1,AEGJM},
which is a much stronger bound than that in Eq.~(\ref{cond1}).

However, for a larger value of $\alpha$, we can still have inflation,
but $V^{\prime}(\phi_0)$ affects the slow roll motion of the inflaton
during the last ${\cal N}_{\rm COBE}$ number of e-foldings.

The total number of e-foldings during the slow roll of $\phi$ from
$\phi_0$ down to $\phi_{\rm end}$ is given by (from now on
$V^{\prime}_0$ and $V^{\prime \prime \prime}_0$ stand for
$V^{\prime}(\phi_0)$ and $V^{\prime \prime \prime}(\phi_0)$)
\beq
\label{tot2} 
{\cal N}_{\rm tot} = {V(\phi_0) \over M^2_{\rm P}}
\int_{\phi_{\rm end}}^{\phi_0} {d\phi \over {V^{\prime}_0 + {1
\over 2} V^{\prime \prime \prime}_0 (\phi - \phi_0)^2}}\,.
\eeq
Note that the two terms in the denominator of the integrand become
equal at $\phi_{\rm eq}$, see Eqs.~(\ref{3rd},\ref{1st}), where
\beq \label{eq}
(\phi_0 - \phi_{\rm eq}) = {1 \over 2} \alpha \phi_0\,.
\eeq

Then we can write
\beq 
\label{tot3} 
{\cal N}_{\rm tot} \simeq {V(\phi_0) \over M^2_{\rm
P}} \Big[\int_{\phi_{\rm eq}}^{\phi_0}{d\phi \over V^{\prime}_0}
+ \int_{\phi_{\rm end}}^{\phi_{\rm eq}}{2 d\phi \over V^{\prime \prime
\prime}_0 (\phi - \phi_0)^2}\Big]\,, 
\eeq
which, after using Eq.~(\ref{eq}), results in~\footnote{Note that
${\cal N}_{\rm tot} \rightarrow \infty$ as $\alpha \rightarrow 0$.
However Eq.~(\ref{tot4}) ceases to hold once $\alpha < (m_{\phi}
\phi^2_0/M^3_{\rm P})^{1/2}$, since as mentioned before, we enter the
self-reproduction regime. The total number of e-foldings in the slow
roll regime is then given by Eq.~(\ref{tot}), which coincides with
Eq.~(\ref{tot4}) when $\alpha \sim (m_{\phi} \phi^2_0/M^3_{\rm
P})^{1/2}$.}
\beq \label{tot4}
{\cal N}_{\rm tot} \simeq {2 \over n (n-1) \alpha} \Big({\phi_0 \over 
M_{\rm P}}\Big)^2\,.
\eeq
It is interesting to note that the two terms on the right-hand side of
Eq.~(\ref{tot3}) gives equal number of e-foldings, i.e. ${\cal N}_{\rm
tot}/2$. The requirement that ${\cal N}_{\rm tot} \geq {\cal N}_{\rm
COBE}$ leads to an upper bound
\beq \label{cond2}
\alpha \ls {2 \over n (n-1){\cal N}_{\rm COBE}} \Big({\phi_0 \over 
M_{\rm P}}\Big)^2\,.
\eeq

Therefore, there exists a window where we can have sufficient number of
e-foldings, but the saddle point results for the spectral index are
not valid any longer. The range of $\alpha$ dictates the trend.
\beq \label{window}
{1 \over n (n-1){\cal N}_{\rm COBE}} \Big({\phi_0 \over 
M_{\rm P}}\Big)^2 < \alpha \ls {2 \over n (n-1){\cal N}_{\rm COBE}} 
\Big({\phi_0 \over M_{\rm P}}\Big)^2\,.
\eeq
Within this window, $\phi_{\rm COBE} > \phi_{\rm eq}$ and, the number
of e-foldings follows
\beq \label{cobe2}
{\cal N}_{\rm COBE} \simeq {V_0 \over M^2_{\rm P}} 
\Big[\int_{\phi_{\rm eq}}^{\phi_{\rm COBE}}{d\phi \over V^{\prime}_0} + 
\int_{\phi_{\rm end}}^{\phi_{\rm eq}}{2 d\phi \over V^{\prime \prime \prime}_0 
(\phi - \phi_0)^2}\Big]\,,
\eeq
instead of Eq.~(\ref{cobe}).

After using Eq.~(\ref{eq}), we find
\beq \label{cobe3}
(\phi_0 - \phi_{\rm COBE}) \simeq \alpha \phi_0 
\Big[1 - {n (n-1)\alpha \over 2} 
\Big({M_{\rm P} \over \phi_0}\Big)^2 {\cal N}_{\rm COBE}\Big]\,.
\eeq
Eq.~(\ref{tilt}) then results in (note that $V^{\prime \prime}(\phi) \approx 
V^{\prime \prime \prime}_0 (\phi - \phi_0)$)
\beq \label{tilt2}
n_s \simeq 1 - 8 n (n-1) \alpha \Big({M_{\rm P} \over \phi_0}\Big)^2 \Big[1 - 
{n (n-1)\alpha \over 2} \Big({M_{\rm P} \over \phi_0}\Big)^2 
{\cal N}_{\rm COBE}\Big]\,.
\eeq
When $\alpha$ saturates the lower bound of the inequality in
Eq.~(\ref{window}), we recover our earlier result, $n_s \simeq 1 -
4/{\cal N}_{\rm COBE} \simeq 0.92$.

On the other hand, when the upper bound of the inequality in
Eq.~(\ref{window}) is saturated, we find $n_s \simeq 1$. This
particular value of $n_s \rightarrow 1$, can be easily understood as,
$\phi_{\rm COBE} \rightarrow \phi_0$, in which case, $\eta \rightarrow
0$.  Therefore the spectral tilt is virtually scale invariant.

Hence within the window we have 
\beq \label{tilt3}
0.92 \leq n_s \leq 1\,.
\eeq

One comment is in order at this point. As pointed out earlier, for the
values of $\alpha$ given in Eq.~(\ref{window}) there will be no
self-reproduction regime in the immediate vicinity of
$\phi_0$. Therefore we actually have slow roll inflation within the
interval $[\phi_e,2\phi_0 - \phi_e]$, where $\phi_e$ is given in
Eq.~(\ref{end}). Note that $V^{\prime}(\phi-\phi_0)$ and $V^{\prime
\prime}(\phi - \phi_0)$ are even and odd respectively near $\phi_0$,
see Eqs.~(\ref{sadapr},\ref{1st}).  This implies that the slope of
potential is symmetric around $\phi_0$, hence there is the same number
of e-foldings in each of the half intervals $[\phi_e,\phi_0]$ and
$[2\phi_0-\phi_e,\phi_0]$. 

One can then consider a situation where $\phi_0 < \phi_{\rm COBE} \leq
2\phi_0-\phi_e$. In the extreme case the total number of e-foldings in
the interval $[\phi_e,2\phi_0-\phi_e]$ is ${\cal N}_{\rm COBE}$
(${\cal N}_{\rm COBE}/2$ in each half interval). This will slightly
increase the upper bound on $\alpha$ obtained in
Eq.~(\ref{window}). Note however that $V^{\prime \prime}(\phi) > 0$,
thus $\eta > 0$, for $\phi > \phi_0$. This implies that $n_s > 1$ if
$\phi_{\rm COBE} > \phi_0$, which is ruled out by the latest WMAP
results~\cite{WMAP3}. For this reason we have only considered the case
where $\phi_{\rm COBE} \leq \phi_0$, which will result in $n_s \leq 1$
(albeit $> 0.92$).

An important point is to note that the fine-tuning in $\alpha$ does
not get any worse if we require that the spectral tilt be closer to
$1$. Eq.~(\ref{cond1}), which implies a bound $\alpha \leq 10^{-9}$ in
order for the saddle point calculations remain valid~\cite{AEGJM}, see
also~\cite{LYTH1} for the discussion on fine tuning. According to
Eq.~(\ref{window}), a larger $n_s$ is found if, $10^{-9} < \alpha \leq
2 \times 10^{-9}$. If any, we have an improvement by a factor of $2$
in $\alpha$.

However, the running of the spectral tilt is slightly different, it
depends on $V'$ and $V^{'''}$ also. The two limiting regimes for
$\alpha$ gives,
\begin{eqnarray}
-\frac{16}{{\cal N}^2_{\rm COBE}}\ls \frac{d\,n_{s}}{d\ln k}\leq
-\frac{4}{{\cal N}_{\rm COBE}^2}\,.
\end{eqnarray}
For a lower limit of $\alpha$, see Eq.~(\ref{window}), we recover our
earlier result, Eq.~(\ref{runtilt}), but for the upper limit, we get a
slightly strong running by virtue of $V'(\phi_0)>V^{'''}(\phi_0)$.
However for ${\cal N}_{\rm COBE} \sim 50$, the running of the spectral
tilt remains very small, i.e. $d\,n_s/d\,{\rm ln}k \sim -0.0064$.

Finally, let us see what happens to the amplitude of the scalar
perturbations within the window in Eq.~(\ref{window}). When the upper
bound of the inequality is saturated, we find
\beq \label{ampl2}
\delta _H \simeq \frac{1}{5\pi} \sqrt{\frac{1}{24}n(n-1)}(n-2) 
\Big({m_\phi M_{\rm P} \over \phi_0^2}\Big) {\cal N}^2_{\rm COBE}.
\eeq
This is smaller by a factor of $4$ compared to the saddle point case,
see Eq.~(\ref{ampl}), which is valid when the lower bound of the
inequality in Eq.~(\ref{window}) is saturated. Therefore within the
allowed window the correct amplitude is obtained with $m_{\phi}$
within the TeV scale range.

%%%%%%%%%%%%%%%%%%%%%%%%%%%%%%%%%%%%%%%%%%%%%%%%%%%%%%%%%%%%%%%%%%%%%%%%%%%%%%%
\section{Conclusion}

MSSM inflation provides interesting possibility that the spectral tilt
can vary in between $0.92\leq n_s \leq 1$ depending on how flat the
potential is in the vicinity of the saddle point. Interestingly, the
upper bound on the spectral tilt $n_s\approx 1$ is found when the
total number of e-foldings is saturated; ${\cal N}_{tot}\approx {\cal
N}_{\rm COBE}$, while the lower limit $n_s \simeq 0.92$ depicts the
robustness of the saddle point inflation. The running of the spectral
tilt remains weak (and well below the observational limits) in both
cases.  The amplitude of perturbations, as well as the fine-tuning of
parameters, remain practically unchanged.

%%%%%%%%%%%%%%
\section{Acknowledgments}
           
We are thankful to our collaborators, Kari Enqvist, Juan
Garcia-Bellido, Asko Jokinen, Alex Kusenko, and also Justin Khoury,
for discussions.  We would particularly like to thank Asko Jokinen for
mentioning some subtle issues on the running of the spectral tilt.
The research of RA is supported by Perimeter Institute for Theoretical
Physics. Research at Perimetre Institute is supported in part by the
Government of Canada through NSERC and by the province of Ontario
through MEDT. The research of AM is partly supported by the European
Union through Marie Curie Research and Training Network
``UNIVERSENET'' (MRTN-CT-2006-035863).

%%%%%%%%%%%%%%%%%%%%%%%%%%%%%%%%%%%%%%%%%%%%%%%%%%%%%%%%%%%%%%%

\end{document}